\documentclass[12pt,iopart,a4paper]{article}
\usepackage{amssymb}
\usepackage{amsfonts}
\usepackage{amsmath}
\usepackage{hyperref}
\usepackage[square,numbers,sort&compress,comma]{natbib}

\hypersetup{colorlinks=true, linkcolor=blue, citecolor=blue, urlcolor=blue}
\input{tcilatex}
\begin{document}

\author{M. S. Bieniek, P. G. C. Almeida, and M. S. Benilov$^{a}$ \\
%EndAName
Departamento de F\'{\i}sica, Faculdade de Ci\^{e}ncias Exatas e da
Engenharia,\\
Universidade da Madeira, Largo do Munic\'{\i}pio, 9000 Funchal, Portugal,
and\smallskip \\
Instituto de Plasmas e Fus\~{a}o Nuclear, Instituto Superior T\'{e}cnico,\\
Universidade de Lisboa, 1041 Lisboa, Portugal\smallskip \\
$^{a}$ email \textit{benilov@uma.pt}}
\title{Self-consistent modelling of self-organized patterns of spots on
anodes of DC\ glow discharges}
\date{}
\maketitle

\begin{abstract}
Self-organized patterns of spots on a flat metallic anode in a cylindrical
glow discharge tube are simulated self-consistently. A standard model of
glow discharges is used, comprising conservation and transport equations for
a single species of ion and electrons, written with the use of the
drift-diffusion and local-field approximations, and the Poisson equation.
Only processes in the near-anode region are considered and the computation
domain is the region between the anode and the discharge column. Multiple
solutions, existing in the same range of discharge current and describing
modes with and without anode spots, are computed for the first time. A
reversal of the local anode current density in the middle of each of the
spots was found, i.e.\ mini-cathodes are formed inside the spots or, as one
could say, anode spots operate as a unipolar glow discharge. The solutions
do not fit into the conventional pattern of self-organization in bistable
nonlinear dissipative systems; e.g. the modes are not joined by bifurcations.
\end{abstract}

\section{Introduction}

\label{Introduction}

For more than a century now beautiful patterns on anodes of DC\ glow
discharges have been observed \cite{Lehmann1902, Mackay1920, Thomas1930,
Rubens1940, Emeleus1982, Mueller1988a, Maszl2011, Arkhipenko2013}. The
patterns are of significant theoretical interest in themselves, and also
because of their connection with various types of double layer structures
and the so-called plasma balls, which have been studied at low gas pressures 
\cite%
{Ivan2005b,Ivan2005c,Charles2007,Baalrud2009,Scheiner2015,Chauhan2016,Scheiner2016,Scheiner2017}%
. The patterns are interesting also from the point of view of their
technological applications. As examples, we note that single spots have been
utilized as ion sources with a stable and high-current ion beam extraction 
\cite{Park2012}, and recently it has been shown that self-organized patterns
on liquid anodes of atmospheric pressure glow microdischarges can reveal a
nontrivial cancer-inhibiting capability \cite{Chen2017b}.

Patterns of spots have been calculated by means of a phenomenological
approach based on the general trends of self-organization in \cite%
{Mueller1988a}. Numerical simulations have been performed which revealed
current density stripes on anodes \cite{Islamov2001}, a circular spot \cite%
{Islamov1998}, and a circular spot surrounded by a ring \cite{Islamov1998}.
A theoretical analysis, and experimental investigation, of the anode layer
region performed in \cite{Akishev2014b} indicates that instabilities found
in the regions' so-called subnormal regime are a precursor for the formation
of anode spots. Also in \cite{Akishev2014b}, the influence that the spots
have on the homogeneity of the plasma column is investigated.

Self-organized arrangements of spots and patterns on cathodes of DC arc and
glow discharges have been understood and systematically described in terms
of multiple steady-state solutions, which exist in conventional models of
glow discharge over the same range of discharge current and describe modes
associated with different cathode spots and cathode spot patterns; e.g., 
\cite{Benilov2014} and references therein. We hypothesize that the same
approach is applicable to spots and spot patterns on anodes of DC glow
discharges. In other words, we postulate that spots and spot patterns on
anodes of DC glow discharges can be described by a new class of solutions,
that exist in conventional models of glow discharges, alongside the solution
associated with the spotless mode of current transfer. In this work we prove
this hypothesis. Two solutions, as examples, are computed over the same,
wide, range of current. One solution describes an axially symmetric diffuse,
or spotless, mode, and the other solution describes a three-dimensional mode
with azimuthal periodicity comprising a self-organized pattern of 8 anode
spots.

The outline of the paper is as follows. The model is described in section %
\ref{Model and Numerics}. In section \ref{Results}, results of the modelling
are given and discussed. Conclusions are drawn in section \ref{Results
copy(1)}.

\section{The model}

\label{Model and Numerics}

Consider a cylindrical DC glow discharge tube that is long enough that the
effect of the electrodes become obviated in the column. In the column the
density of charged species and electric field are independent of the axial
coordinate. This invariance allows us to choose an asymptotically accurate
set of boundary conditions on a domain that contains only the region from
the anode to the column. The computation domain is adequate for an
investigation of anode spots, or patterns of spots, appearing as a result of
processes of plasma-anode interaction only.

The simplest model of a glow discharge is used, which is well-known but
briefly summarized here for completeness. It comprises equations for
conservation of electrons and a single ion species, the transport equations,
written in the drift-diffusion approximation, and Poisson's equation:%
\begin{eqnarray}
\frac{\partial n_{i}}{\partial t}+\nabla \cdot \mathbf{J}_{i}
&=&n_{e}\,\alpha \,\mu _{e}\,E-\beta \,n_{e}\,n_{i},\;\;\;\mathbf{J}%
_{i}=-D_{i}\,\nabla n_{i}-n_{i}\,\mu _{i}\,\nabla \varphi ,  \notag \\
\frac{\partial n_{e}}{\partial t}+\nabla \cdot \mathbf{J}_{e}
&=&n_{e}\,\alpha \,\mu _{e}\,E-\beta \,n_{e}\,n_{i},\;\;\;\mathbf{J}%
_{e}=-D_{e}\,\nabla n_{e}+n_{e}\,\mu _{e}\,\nabla \varphi ,  \notag \\
\varepsilon _{0}\,\nabla ^{2}\varphi &=&-e\,\left( n_{i}-n_{e}\right) .
\label{1}
\end{eqnarray}%
\newline
Here $n_{i}$, $n_{e}$, $\mathbf{J}_{i}$, $\mathbf{J}_{e}$, $D_{i}$, $D_{e}$, 
$\mu _{i}$, and $\mu _{e}$ are number densities, densities of transport
fluxes, diffusion coefficients, and mobilities of the ions and electrons,
respectively; $\alpha $ is Townsend's ionization coefficient; $\beta $ is
coefficient of dissociative recombination; $\varphi $ is electrostatic
potential, $E=\left\vert \nabla \varphi \right\vert $ is electric field
strength; $\varepsilon _{0}$ is permittivity of free space; and $e$ is
elementary charge. The local-field approximation is employed, i.e.,\
electron transport and kinetic coefficients are assumed to depend on the
local electric field only.

Let us introduce cylindrical coordinates $\left( r,\phi ,z\right) $ with the
longitudinal axis in line with the axis of the discharge tube. The
computation domain is a cylinder \newline
$\left\{ 0\leq r\leq R,0\leq \phi \leq 2\pi ,0\leq z\leq h\right\} $, where $%
R$ is the tube radius and the boundary $z=h$ is positioned in the discharge
column.

Standard boundary conditions are used for the lateral dielectric\ wall, $r=R$%
,\ describing absorption of ions and electrons, and electrical insulation: 
\begin{equation}
J_{in}=\sqrt{\frac{8k_{B}T_{i}}{\pi m_{i}}}\frac{n_{i}}{4},\;J_{en}\mathbf{=}%
\sqrt{\frac{8k_{B}T_{e}}{\pi m_{e}}}\frac{n_{e}}{2},\ J_{in}-J_{en}=0.
\label{2}
\end{equation}%
Here subscript $n$ represents the projection of the corresponding vector
along $n$ the normal directed outside the computation domain, $k_{B}$ is
Boltzmann's constant, $T_{i}$ and $T_{e}$ are ion and electron temperatures
(known parameters), $m_{i}$ and $m_{e}$ are the ion and electron masses.
When a time-dependent solver is used, the last condition in equation (\ref{2}%
) is replaced by the following boundary condition%
\begin{equation}
-\varepsilon _{0}\frac{\partial \varphi }{\partial n}=\rho _{s}\text{, }%
\frac{\partial \rho _{s}}{\partial t}=e\left( J_{in}-J_{en}\right) \text{,}
\label{accum}
\end{equation}%
which describes surface charge accumulation; here $\varepsilon _{0}$ is
permittivity of free space and $\rho _{s}$ is surface charge density. If a
steady-state has been reached, these conditions are equivalent to the last
condition in equation (\ref{2}).

Boundary conditions at the anode surface ($z=0$) are 
\begin{equation}
J_{in}=\sqrt{\frac{8k_{B}T_{i}}{\pi m_{i}}}\frac{n_{i}}{4}\text{$,$ }J_{en}%
\mathbf{=}\sqrt{\frac{8k_{B}T_{e}}{\pi m_{e}}}\frac{n_{e}}{2}-\delta \gamma
J_{in},\text{ }\varphi =0.  \label{3}
\end{equation}%
The conditions for the ions and the electrons are similar to the ones for
the dielectric wall (\ref{2}), except for the second term on the rhs of the
boundary condition for the electrons (the second equation in (\ref{3})).
This term describes secondary electron emission, which may become relevant
if the local electric field is directed from the plasma to the anode. A
parameter $\delta $ is introduced which is $1$ if the local electric field
is directed to the anode, and $0$ otherwise. Note that the choice of which
secondary electron emission coefficient, $\gamma $, to use was not clear as
the anode sheath voltage and, consequently, the energy of incident ions are
small. In any case, this term produces a small effect even for $\gamma $ of
order unity, since its magnitude for comparable$\ n_{i}$ and $n_{e}$ is of
the order of $\gamma \sqrt{m_{e}T_{i}/\left( m_{i}T_{e}\right) }$ with
respect to the first term on the rhs of the second equation in (\ref{3}).
The third condition in (\ref{3}) defines the zero of potential.

The boundary $z=h$ is positioned in the discharge column, where the charged
species densities are independent of $z$ and the axial electric field is
constant (independent of $r,\phi ,z$):

\begin{equation}
\frac{\partial n_{i}}{\partial n}=0,\frac{\partial n_{e}}{\partial n}=0,\ \ 
\frac{\partial \varphi }{\partial n}=E_{z}\text{. }  \label{4}
\end{equation}%
Here $E_{z}$ is the axial electric field; a given parameter which may be
chosen to ensure desired values of the discharge current $I$. The parameter $%
h$ has to be large enough so the conditions (\ref{4}) are satisfied not just
at the boundary $z=h$, but also in a region adjacent to the boundary; in
other words, $h$ has to be larger than the thickness of the near-anode
region.

We hypothesize that the problem (\ref{1})-(\ref{4}) admits an axially
symmetric (2D) steady-state solution, describing a spotless, or diffuse,
mode of current transfer to the anode, and three-dimensional steady-state
solutions, presumably describing modes with patterns of spots. By analogy
with computed spot patterns on cathodes of DC glow discharges, and in
qualitative agreement with experimental results on anode spot patterns, we
assume that the 3D solutions are periodic in $\phi $ with the period $2\pi
/n $, where $n=1,2,3,\dots $, then it is sufficient to limit the computation
domain to a half-period of the desired 3D solution: $0\leq \phi \leq \pi /n$%
. Boundary conditions at $\phi =0$ and $\phi =\pi /n$ are zero normal
derivatives,%
\begin{equation}
\frac{\partial n_{i}}{\partial n}=0,\frac{\partial n_{e}}{\partial n}=0,\ 
\frac{\partial \varphi }{\partial n}=0,  \label{5}
\end{equation}%
so that $\phi =0$ and $\phi =\pi /n$ represent planes of symmetry of the
solution considered.

Results reported in this work refer to a discharge in helium under the
pressure of $5\,\mathrm{Torr}$. The (only) ionic species considered is $%
\mathrm{He}_{2}^{+}$. The transport and kinetic coefficients are the same as
in \cite{Almeida2013}. The discharge tube radius is $R=0.5\unit{mm}$ and the
height of the computation domain is $h=5\unit{mm}$. It is set that $\gamma
=0.03$, $T_{e}=1\unit{eV}$, and $T_{i}=300\unit{K}$.

The modelling was performed in COMSOL Multiphysics. Both the steady-state
and time-dependent forms of problem (\ref{1})-(\ref{5}) have been solved.
The Plasma Module with\ a stationary solver, and a time-dependent solver,
have been employed. The Plasma Module was adapted so that it could be used
in combination with a stationary solver and supplemented with a
residual-based stabilization method.

\section{Results and discussion}

\label{Results}

One of the computed solutions reported in this paper is 2D and describes the
spotless mode. As an example, a 3D mode with $n=8$ is also reported, and it
describes a mode with eight spots. Note that the relatively high value of $n$
permits a relatively small computation domain and thus reduces the required
RAM and computation time.

The 2D solution was computed in a standard way by means of a stationary
solver. It has been found in this work that 3D solutions do not bifurcate
from the 2D solution, in contrast to solutions describing cathodic spots and
patterns of spots in arc and glow cathodes, which do bifurcate from a
fundamental (generally 2D) solution. Therefore the approach developed for
the systematic computation of multiple solutions describing spots and
patterns on cathodes of arc and DC glow discharges \cite{Benilov2014} could
not be used. To find the 3D solution reported in this work, we first solved
the 1D axially symmetric and steady-state form of the problem (\ref{1})-(\ref%
{2}), describing the discharge column. (Analytical solutions of this 1D
problem for the limiting cases corresponding to free-fall and ambipolar
diffusion \cite{Franklin1976} and a recombination-dominated discharge were
used to validate the code.) In order to obtain the 3D solution, a solution
of the 1D problem governing the column for the discharge current $I=10\unit{%
mA}$ was introduced as an initial condition for the time-dependent solver,
the one solving the time-dependent form of the problem (\ref{1})-(\ref{5})
including surface charge accumulation at the dielectric wall, equation (\ref%
{accum}). The computations have been performed with the value of $E_{z}$,
the input parameter describing the axial electric field at the column
boundary, corresponding to the $I=30\unit{mA}$, and not to $10\unit{mA}$.\
The time-dependent solver was ran; the mismatch in $E_{z}$ introduced a
perturbation to the system that resulted in an evolution to a 3D
time-independent solution to the problem. The stationary solver was then
used to compute the 3D solution in a wide range of current.

\subsection{Current-voltage characteristics of the anode region}

Consider the potential distribution in the discharge column, $\varphi _{c}$,
(which is axially symmetric), 
\begin{equation}
\varphi _{c}\left( r,z\right) =-(z-h)E_{z}+\varphi _{h}\left( r\right) ,
\label{6}
\end{equation}%
where $\varphi _{h}\left( r\right) $ is the distribution of potential at the
computational boundary, $z=h$. We define the near-anode voltage drop as the
difference between the potential at the anode (equal to zero), and the
potential that is obtained by extrapolation of the column solution (\ref{6})
to the anode ($z=0$):

\begin{equation}
U=-hE_{z}-\varphi _{h}\left( r\right) .  \label{7}
\end{equation}

Note the second term on the rhs of this definition depends on $r$. In order
to remove the dependence on $r$ and find an integral characteristic, one
has, for example, to evaluate the rhs of equation (\ref{7}) on the axis, or
edge, of the discharge tube, or take an average value over the cross
section. However, whatever choice is made is irrelevant in so far as a
graphical representation of multiple solutions is concerned: different
solutions with the same discharge current will coincide in the column to the
accuracy of a shift of potential by a constant. Therefore, whatever way is
chosen to evaluate the rhs of equation (\ref{7}), as long as it is the same
for different solutions, the difference in $U$ between the different
solutions will be equal. We indicate for definiteness that in this work the
rhs of equation (\ref{7}) is evaluated on the axis.

Shown in Figure 1 are near-anode current-voltage characteristics (CVCs) of
two solutions, existing in the same range of current. One solution describes
a 3D mode that is azimuthally periodic, the other a 2D mode that is axially
symmetric. The 3D mode has a negative value of the near-anode voltage in the
range of the computed current, while the 2D mode has a positive value of the
near-anode voltage in the range of the computed current.\ 

It is of interest to compare the CVCs\ in figure 1 to the computed CVCs for
DC\ glow discharges with self-organized cathode spots (e.g. figure 3 of \cite%
{Bieniek2016}). In the case of the cathode spots there is an N-shaped CVC
corresponding to the 2D solution, with the CVC corresponding to the 3D
solutions\ branching off from near the falling section of the CVC\ of the 2D
solution; as per the general pattern of self-organization in bistable
nonlinear dissipative systems with a positive feedback. The CVCs shown in
figure 1 are very different: no pronounced N-shape CVC was computed for the
2D solution, and no bifurcations were found in a wide current range.

\subsection{Anode spot structure}

The electron number density\ on the surface of the anode for the 3D mode at $%
10\unit{mA}$ is shown in figure 2. Electron density is organized in an
azimuthally periodic pattern of spots. The pattern is similar to that
observed experimentally \cite[Figure 1]{Mackay1920}.

Densities of ions and electrons are shown in figure 3 in the plane of
symmetry $\phi =0$ (a longitudinal cross section passing through the tube
axis and the centre of a spot) for $I=1\unit{mA}$. One can see that the
electron and ion densities are distributed in a similar way to the so-called
fireballs observed experimentally in \cite{Baalrud2009}.

The distribution of the charged particle densities along the axial direction
from the centre of a spot to the end of the calculation domain (a 1D plot
with constant $\phi ,r$), for the 3D mode at $I=0.1\unit{mA}$ and $35\unit{mA%
}$, are shown in figure 4. There is a region with $n_{i}>n_{e}$, i.e., an
ion sheath, adjacent to the electrode. The ion densities in the sheath are
of the same order of magnitude for the two discharge currents, while the
densities of the charged particles in the column vary by more than an order
of magnitude. For $I=0.1\unit{mA}$, charge separation is seen also in the
column, which is due to ambipolar diffusion coming into play near the
(absorbing) lateral wall.

The distribution of the electric field for $I=35\unit{mA}$ is shown in
figure 4. The electric field in the ion sheath is two orders of magnitude
greater than that in the quasi-neutral region. The former points towards the
anode, while the latter points away from the anode.

It is seen in figure 3 c) that near the spot, the electrode is biased below
the potential of the adjacent plasma, and it is seen from figure 4 that
quasi-neutral plasma is extended close, up to $1\unit{%
%TCIMACRO{\U{3bc}}%
%BeginExpansion
\mu%
%EndExpansion
m}$, to the electrode. In the spotless mode at the same current, the
electrode potential is above that of the adjacent plasma and the electron
sheath extends $50\unit{%
%TCIMACRO{\U{3bc}}%
%BeginExpansion
\mu%
%EndExpansion
m}$ from the electrode. It is seen from figure 3 c) that the electrode has a
positive bias with respect to the adjacent plasma at large distances from
the spot; skipping for brevity results on the charge particle distribution,
we only note that there is an electron sheath adjacent to the electrode far
from the spot.

\subsection{Near-anode physics}

The distribution of current density and electric field along the anode
surface in the plane of symmetry $\phi =0$ (the longitudinal cross section
that passes through the centre of a spot) is shown in figure 5. Plots are
included for two different discharge currents. The current density has a
large magnitude and is negative inside the spot, and turns positive outside.
The spot behaves like a mini-cathode or, as one could say, operates as a
unipolar glow discharge.

The direction of current density in the plane of symmetry is shown in figure
6. For convenience, the distribution of the ions number density (the same as
in figure 3b) is shown as well. The unipolar glow discharge is clearly seen.

The electric field at the anode in figure 5 is seen to be both negative
inside the spot and positive outside for $35\unit{mA}$; it is negative
everywhere for $1\unit{mA}$. Directions of the electric field at the anode
and of current density from the anode inside and outside the spot are
summarized in table 1. Also shown are corresponding data for the 2D\
spotless mode, where $E_{z}>0$, $j_{z}>0$ for all values of current.\bigskip

\begin{tabular}{|c|c|c|c|}
\hline\hline
Current & Within the spot & Outside the spot & Spotless mode \\ \hline\hline
1 $\unit{mA}$ & 
\begin{tabular}{l}
$E_{z}<0$, $j_{z}<0$ \\ 
(cathode)%
\end{tabular}
& 
\begin{tabular}{l}
$E_{z}<0$, $j_{z}>0$ \\ 
(field reversed-anode)%
\end{tabular}
& 
\begin{tabular}{l}
$E_{z}>0$, $j_{z}>0$ \\ 
(regular anode)%
\end{tabular}
\\ \hline\hline
35 $\unit{mA}$ & 
\begin{tabular}{l}
$E_{z}<0$, $j_{z}<0$ \\ 
(cathode)%
\end{tabular}
& 
\begin{tabular}{l}
$E_{z}>0$, $j_{z}>0$ \\ 
(regular anode)%
\end{tabular}
& 
\begin{tabular}{l}
$E_{z}>0$, $j_{z}>0$ \\ 
(regular anode)%
\end{tabular}
\\ \hline
\end{tabular}

Table 1. Physics of the near-anode region for the 3D\ solution.\newline

\section{Concluding remarks}

\label{Results copy(1)}

For the first time, a self-organized pattern of spots of plasma on an anode
was computed self-consistently. A new class of stationary solutions,
describing anode spots, was found in a conventional DC\ glow discharge
model. The 3D solution was found to exist in the same range of currents as a
2D solution describing a spotless mode of current transfer.

There are similarities between the patterns of anode spots and the patterns
of spots on cathodes of arc and DC glow discharge: both are described by
multiple steady state solutions and reveal azimuthal periodicity. On the
other hand, the spots on the anode are different to the spots on the cathode
in following ways: the solution describing the spotless mode does not
contain a pronounced N-shaped current-voltage characteristic; no
bifurcations were found in a wide range of currents, i.e., the anode spots
were found to exist isolated from the 2D spotless mode. The anode spots are
apparently related to the change in the sign of the near anode voltage.

Inside the spots the anode behaves like a mini-cathode, in that the sign of
the current density and electric field is reversed. In other words, anode
spot operates as a unipolar glow discharge.

The above-described physics is different from the physics revealed by the
recent modelling \cite{Scheiner2016,Scheiner2017} of plasma balls, that form
on a small anode in contact with a low-pressure plasma and sometimes are
termed anode spots. In particular, no double layers were found in the
present modelling. One of the reasons of the difference is that the plasma
balls appears not to be a self-organization phenomenon; note that its size
is bigger than the width of the electrode. The difference in plasma pressure
may contribute as well.

Bombardment on the anode by low-kinetic energy ions occurs inside the spots.
An interesting hypothesis is that the ions incident on the anode contribute
to or are responsible for the cancer-inhibiting effect reported in \cite%
{Chen2017b}.

The modelling reported in this work should not be interpreted as an attempt
to quantitatively describe parameters of anode spots. Merely, the aim was to
prove the possibility of self-consistently describing self-organized anode
spots on the basis of multiple solutions existing in the same range of
discharge currents, which was achieved. The model may be used for
qualitative analysis, and certainly some interesting trends have emerged.
Nevertheless it is well known that a detailed account of the complex plasma
chemistry of a helium discharge, and the nonlocality of electron energy
ought to be included. Another aspect that needs to be improved is a
description of the near-electrode sheath, which is collisionless inside the
spot.

\section{Acknowledgments}

The work was supported by FCT - Funda\c{c}\~{a}o para a Ci\^{e}ncia e a
Tecnologia of Portugal through the project Pest-OE/UID/FIS/50010/2013.

\bibliographystyle{apsrev4-1}
\bibliography{Matt_thesis}

\end{document}